\documentclass[showpacs, %
 reprint,
 amsmath,amssymb,
 aps,
]
{revtex4-1}

\usepackage{graphicx}
\usepackage{dcolumn}
\usepackage{bm}

\begin{document}
\preprint{APS/123-QED}
\title{Theory of unconventional spin states in surfaces with non-Rashba spin-orbit interaction}
\author{Kokin Nakajin$^{1,2}$}
\author{Shuichi Murakami$^{1,2}$}
\affiliation{${}^{1}$Department of Physics, Tokyo Institute of Technology, 2-12-1 Ookayama, Meguro, Tokyo 152-8551, Japan}
\affiliation{${}^{2}$CREST, JST, 4-1-8 Honcho, Kawaguchi, Saitama, 332-0012, Japan}

\begin{abstract}

Surface states in Tl/Si(111)-(1$\times$1) and $\beta$-Bi/Si(111)-($\sqrt{3}\times\sqrt{3}$) show non-Rashba-type spin splitting.
We study spin-transport properties in these surface states.
First we construct tight-binding Hamiltonians for Tl/Si and Bi/Si surfaces, which respect crystallographic symmetries.
As a result, we find specific terms in the Tl/Si surface Hamiltonian responsible for non-Rashba spin splitting. Using this model we calculate current-induced spin polarization in the Tl/Si Hamiltonian in order to see the effect of non-Rashba spin-orbit interaction. We found that the induced spin polarization is in-plane and perpendicular to the current, which is consequently the same with Rashba systems. We find that it follows from crystallographic symmetries. 
Furthermore, we numerically find bound states at the junction between two surface regions which have different signs of the spin-orbit interaction parameters in the Bi/Si system and in the Tl/Si system.
We explain these numerical results with the results of our analytical calculations.

\end{abstract}
\maketitle


\section{Introduction}
The spin-orbit interaction (SOI) in crystals has various forms depending on materials and their symmetries.
In a surface of a metal, one of the typical forms of the SOI is the Rashba SOI $(\vec{\sigma}\times\vec{k})_z$ \cite{0022-3719-17-33-015} and is observed in 
metal surfaces such as in Au(111) surface \cite{lashell1996spin}.
Here the Fermi surfaces consist of two concentric circles, and spins are tangential to the Fermi surfaces. 
This spin splitting effect has been observed on clean noble metal surfaces \cite{PhysRevB.65.033407,PhysRevB.69.241401,PhysRevB.72.045419,PhysRevB.73.195413} and heavy group V elements \cite{PhysRevLett.93.046403,PhysRevLett.96.046411,PhysRevLett.96.046411,PhysRevLett.97.146803}.
On the other hand, bulk inversion asymmetry gives rise to another type of the SOI called 
Dresselhaus SOI \cite{PhysRev.100.580}. The Dresselhaus SOI in two dimensions has the form: $k_x\sigma_x-k_y\sigma_y$ to the linear order in $\vec{k}$. 
Its energy band splitting is similar to that of the Rashba SOI, but 
the spin direction is unlike the Rashba SOI, 
as has been observed in the GaAs(110) surface \cite{PhysRevLett.52.2297}.

Since there are various types of SOI terms depending on systems, we can 
expect rich physics from their interplay.
For example, two-dimensional (2D) systems including both the Rashba and Dresselhaus SOIs with equal magnitude have an anomalously enhanced spin lifetime \cite{PhysRevLett.97.236601}.
Because of nested Fermi surfaces with opposite spins, the spin life time is proposed to be largely enhanced at the nesting wave vector.
This phenomenon was observed in 2D
electron gas \cite{PhysRevLett.98.076604} in semiconductor quantum wells \cite{N458610} and is called persistent spin helix.

Thanks to the low symmetries of material surfaces, 
they sometimes allow even other types of SOI.
For example, non-Rashba-type surface states have been measured in Tl/Si(111)-($1{\times}1$) \cite{PhysRevLett.102.096805} and $\beta$-Bi/Si(111)-($\sqrt{3}\times\sqrt{3}$) surfaces \cite{PhysRevLett.103.156801} by angle-resolved photoemission spectroscopy (ARPES).
On the Tl/Si surface the spin-split states at the $\bar{K}$ point have the spins 
normal to the surface.
In addition, on the Bi/Si surface there is a ``peculiar'' Rashba splitting at the $\bar{K}$ points. It is called peculiar because the $\bar{K}$ points 
are not time-reversal invariant, whereas the conventional Rashba splitting appears only around time-reversal invariant $\vec{k}$ points \cite{0953-8984-21-9-092001,oguti.shisshido}. 
Furthermore, at the $\bar{M}$ point in the Bi/Si surface, the spin texture is not of Rashba type, but is similar to that of Dresselhaus type.
Such unconventional systems may have new spin properties which are yet to be discovered.

In this paper, we theoretically explore new properties due to the SOI in the Tl/Si(111)-($1{\times}1$) \cite{PhysRevLett.102.096805} and $\beta$-Bi/Si(111)-($\sqrt{3}\times\sqrt{3}$)  \cite{PhysRevLett.103.156801} surfaces.
First, we construct effective tight-binding Hamiltonians of the Tl/Si and Bi/Si surfaces.
We verify qualitative agreement for energy bands and spin texture, between the experimental results and our results.
Second, we explore spin properties in the non-Rashba-type system, Tl/Si, 
such as the current-induced spin polarization and the persistent spin helix. We show that a charge current induces spin polarization in the system with non-Rashba-type spin splitting. Although the Hamiltonian contains out-of-plane components, the induced spin is shown to be in-plane.
In addition, when the Fermi energy is controlled in the Tl/Si surface so that there are two nested carrier pockets with opposite spins, 
we theoretically show that spin helix states can be realized  at a nesting wave vector.

Additionally, we study a junction between two surface regions which have different signs of the SOI parameters in the Bi/Si  and  the Tl/Si systems.
It is motivated by the related work in topological insulators (TIs) \cite{PhysRevLett.107.166805}. The junction of TIs with different sizes of the SOI is shown to exhibit
a novel refraction phenomenon, and in addition it was shown that there exist topologically-protected gapless interface states between two TIs when the SOI of the two TIs have opposite signs \cite{PhysRevLett.107.166805}.
Because TIs and Rashba systems have the same form of the SOI, we expect a similar behavior in Rashba systems.
Motivated by this work, we study a junction between two surface regions with non-Rashba SOI of opposite signs. We see that in some cases the junctions support bound states, with its spin directions different from the bulk states.


\section{Construction of effective models for the non-Rashba systems}

In this section we construct effective tight-binding models for two types of non-Rashba systems.
These models are intended to be minimal models sharing the same symmetry properties as the original systems. Hence they do not necessarily reproduce the band structures of the original materials quantitatively.

\subsection{Tl/Si(111) surface}
\begin{figure}[b]
 \begin{center}
  \includegraphics[width=80mm]{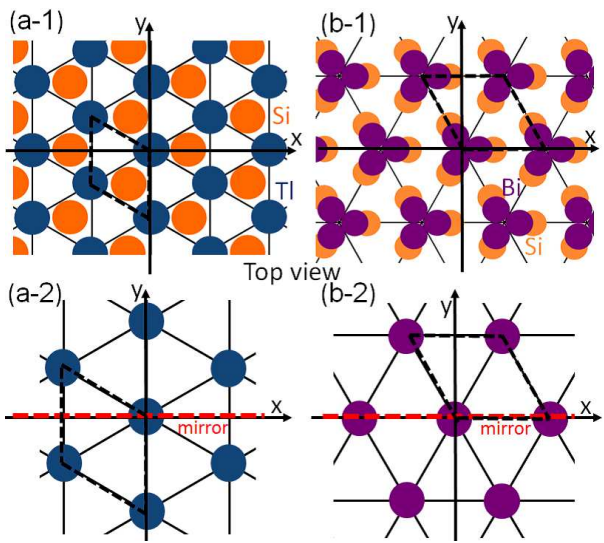}
 \end{center}
 \caption{(Color online)\ (a-1) Schematic illustration of the Si(111)-($1{\times}1$) surface with  adsorption of one monolayer of Tl. The orange balls indicate the topmost Si atoms and the blue balls indicate the Tl atoms.
(b-1) Schematic illustration of the Si(111)-($\sqrt{3}\times\sqrt{3}$) surface
 with adsorption of one monolayer of Bi. The purple balls indicate the Bi atoms.
In (a) and (b) the black dashed lines indicate the unit cell of each surface.
(a-2) and (b-2) represent the triangular lattice used for the tight-binding models. 
In (a-2) the blue balls denote the Tl atoms, while in (b-2) each purple ball denotes
a set of three Bi atoms. The red dashed lines indicate the mirror planes ($xz$ plane).
}  \label{fig:0}
\end{figure}
We construct an effective tight-binding Hamiltonian for the Tl/Si surface [Fig.~\ref{fig:0}(a-1).].
We consider the model on the triangular lattice, which represents the Tl atoms as 
shown in Fig.~\ref{fig:0}(a-2).  
We retain only the nearest-neighbor hoppings, and 
the $x$ and $y$ directions are defined as in Fig.~\ref{fig:0}(a-2).
By taking into account the symmetries of the Tl/Si crystal, such as three-fold rotational symmetry $C_{3z}$, time-reversal symmetry, and mirror symmetry with respect to the $xz$ plane $M_{xz}$ [see Fig.~\ref{fig:0}(a-2)], the Hamiltonian is represented as 
\begin{eqnarray}
H=\sum_{\langle {i},{j}\rangle}C_{i}^\dag
[t+i\lambda_x(\vec{\sigma}\times\vec{d}_{ij})_z+i\lambda_z\xi_{ij}{\sigma}_z
]C_{j},\label{eqn:1}
\end{eqnarray}
where $\vec{\sigma}$ is the vector of the Pauli matrices, $\vec{d}_{ij}$ is the vector from site $i$ to $j$, 
and $C_i$ is an annihilation operator for an electron at the $i$-th site in the triangular lattice. $\xi_{ij}$ takes the values $\pm 1$ depending on the hopping directions: 
 $\xi_{ij}$ is $\pm 1$ for $\theta =(4n\mp 1)\pi/6$ ($n$: integer), where $\theta$ is an angle between the hopping vector and the $+x$ direction.
The second term of the Hamiltonian involves the in plane spin perpendicular to the hopping direction. The third term involves the spin perpendicular to the plane (${\parallel}z$), and 
such term does not exist in Rashba systems.
It causes non-Rashba spin splitting with spin polarization perpendicular to the crystal surface 
at the $\bar{K}$, $\bar{K'}$ points, as we see in the following.

\begin{figure}[h]
 \begin{center}
  \includegraphics[width=80mm]{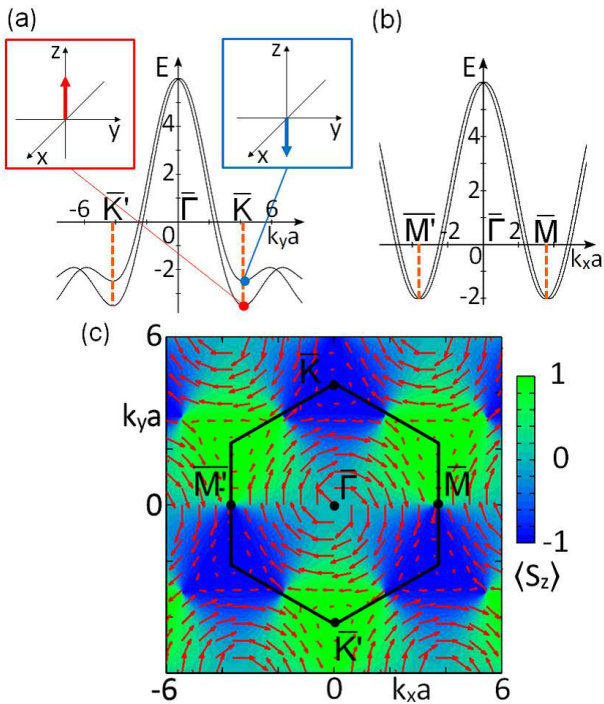}
 \end{center}
 \caption{(Color online) \ Energy bands for effective surface Hamiltonian of Tl/Si. Band structures are plotted along (a) the $\bar{K}$-$\bar{\Gamma}$-$\bar{K'}$ and (b) the $\bar{M}$-$\bar{\Gamma}$-$\bar{M'}$ directions. Spins at $\bar{K}$ and $\bar{K}'$ are
shown as
blue and red arrows in (a). (c) Spin configuration of the upper band ($\eta=1$) in the first Brillouin zone. Arrows represent the in-plane spin direction (${\langle}S_x\rangle,{\langle}S_y\rangle$) and the color 
represents the out-of-plane spin component ${\langle}S_z\rangle$.
The black hexagon represents the surface Brillouin zone.
The parameters are set as $t=1$, $\lambda_{x,z}=0.1$.
}
 \label{fig:1}
\end{figure}

The Hamiltonian (\ref{eqn:1}) is rewritten into a matrix form
\begin{eqnarray}
H(\vec{k})=d_0+d_1\sigma _x+d_2\sigma _y+d_3\sigma _z, \label{eqn:2}
\end{eqnarray}
where
\begin{eqnarray}
d_0&=&t(2\cos{2Y}+4\cos{X}\cos{Y}), \label{eqn:d-1}\\
d_1&=&\lambda_x(2\sin{2Y}+2\cos{X}\sin{Y}),\label{eqn:d-2}\\
d_2&=&-\lambda_x(2\sqrt{3}\sin{X}\cos{Y}),\label{eqn:d-3}\\
d_3&=&\lambda_z(2\sin{2Y}-4\cos{X}\sin{Y}),\label{eqn:d-4}
\end{eqnarray}
where $X=\frac{\sqrt{3}}{2}k_xa$, $Y=\frac{1}{2}k_ya$, and $a$ is the lattice constant.
Its eigenvalue and spin direction are given by
\begin{eqnarray}
E(\vec{k})&=&d_0+\eta\sqrt{{d_1^2+d_2^2+d_3^2}},\label{eqn:2-1}\\
{\langle}\vec{s}{\rangle}&=&\eta\hat{d},\label{eqn:2-2}
\end{eqnarray}
where
$\eta=\pm1$, $\hat{d}$ is a unit vector along $\vec{d}=(d_1,d_2,d_3)$.
The surface Brillouin zone is a hexagon, with its corners 
$\bar{K}(0,\frac{4\pi}{3a})$, $\bar{K'}(0,-\frac{4\pi}{3a})$. As an example, we numerically
calculate the band structure for $t=1$, $\lambda_{x,z}=0.1$. 
At $\bar{K}$ and $\bar{K'}$ points $d_1=d_2=0$, and the Hamiltonian reduces to $d_0+d_3\sigma _z$.
Hence the energy bands around the $\bar{K}$, $\bar{K'}$ points have non-Rashba splitting, with the spin directions perpendicular to the crystal surface [Fig.~\ref{fig:1}(a)].
On the other hand, at $\bar{M}(\frac{2\pi}{\sqrt{3}a},0)$, $\bar{M'}(-\frac{2\pi}{\sqrt{3}a},0)$ and $\bar{\Gamma}(0,0)$, $\vec{d}$ becomes zero, and the two bands are degenerate. 
Around $\bar{M}$, $\bar{M'}$ and $\bar{\Gamma}$ points, the vector $\vec{d}$ forms a vortex, and consequently the bands have Rashba splitting (Fig.~\ref{fig:1}(b)).
These features of our results qualitatively agree with the experimental results for the Tl/Si(111) surface \cite{PhysRevLett.102.096805}.
The spin distribution for the present model in the first Brillouin zone is shown in Fig.~\ref{fig:1}(c). This is in agreement with the above considerations. 

Here, we note a previous work on theoretical calculation for the Tl/Si crystal by a different method \cite{PhysRevB.80.241304}. It is based on a four-band effective model on the honeycomb lattice representing the Tl atoms and Si atoms in the first layer. The calculated 
spin configuration qualitatively agrees with our results. Compared with the four-band model Hamiltonian, our two-band model is simpler and useful for analytic calculations and investigations of new phenomena.


\subsection{$\beta$-Bi/Si(111) surface}

\begin{figure}[htb]
 \begin{center}
  \includegraphics[width=80mm]{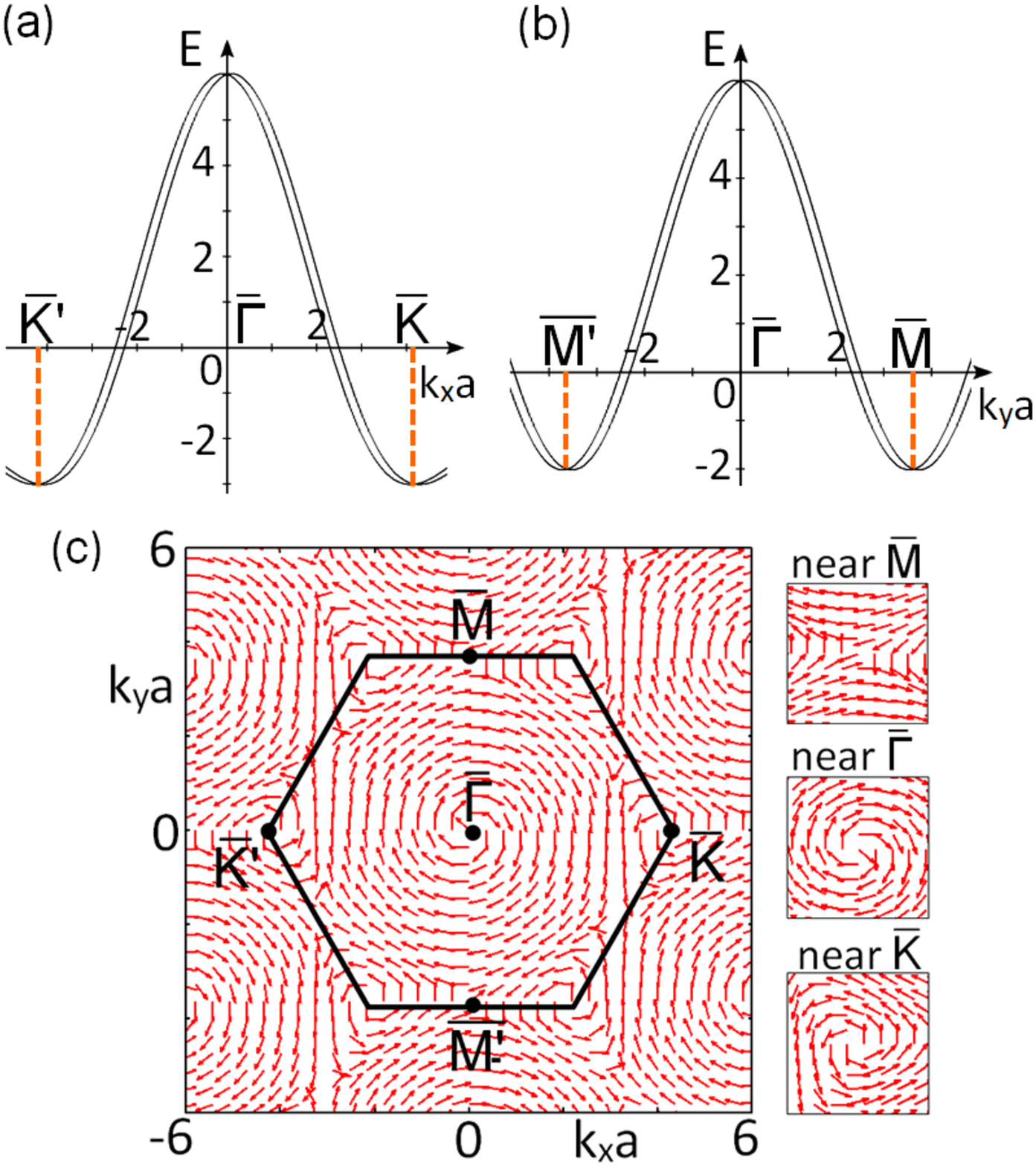}
 \end{center}
 \caption{(Color online)\ Energy bands and spins for effective surface Hamiltonian of the Bi/Si. Band structures are plotted along (a) the $\bar{K}$-$\bar{\Gamma}$-$\bar{K'}$ and (b) the $\bar{M}$-$\bar{\Gamma}$-$\bar{M'}$ directions.
(c) Spin polarization for the upper band ($\eta=1$). Each arrow represents the in-plane spin direction (${\langle}S_x\rangle,{\langle}S_y\rangle$). 
The black hexagon represents the surface Brillouin zone.
 The parameters are set as $t=1$, $\lambda_{y}=0.1$.}
 \label{fig:3}
\end{figure}

Next, we take the same procedure for the Bi/Si crystal surface. 
We construct an effective tight-binding Hamiltonian on the triangular lattice, representing the states close to the Fermi energy.
Symmetries of the Bi/Si crystal to be considered are the $C_{3z}$ symmetry, the time-reversal symmetry, and the mirror symmetry $M_{xz}$ [see Fig.~\ref{fig:0}(b-2)]. 
The symmetry properties might look the same with Tl/Si(111) but it is not the case. As can be seen from Fig.~\ref{fig:0}(a-2) and (b-2), the relative positions of mirror planes in the triangular lattice are different, and it brings about different restrictions for the effective Hamiltonian.
The resulting effective model with only the nearest-neighbor hoppings is written as
\begin{eqnarray}
H=\sum_{\langle i,j \rangle}C_{i}^\dag
[t+i\lambda_y(\vec{\sigma}\times\vec{d}_{ij})_z
]C_{j}.\label{eqn:3}
\end{eqnarray}
This Hamiltonian does not contain a term similar to the third term in Eq.~(\ref{eqn:1}).
The Hamiltonian matrix can be written as
\begin{eqnarray}
H(\vec{k})=d_0+d_1\sigma _x+d_2\sigma_y, \label{eqn:a-1}
\end{eqnarray}
where
\begin{eqnarray}
d_0&=&t(2\cos{2X}+4\cos{X}\cos{Y}), \label{eqn:d-5}\\
d_1&=&\lambda_y(2\sqrt{3}\cos{X}\sin{Y}),\label{eqn:d-6}\\
d_2&=&-\lambda_y(2\sin{X}\cos{Y}+2\sin{2X}),\label{d-7}
\end{eqnarray}
where $X=\frac{1}{2}k_xa$, $Y=\frac{\sqrt{3}}{2}k_ya$, and $a$ is the lattice constant.
As an example, we calculate the band structure for $t=1$ and $\lambda_y=0.1$, shown in 
Fig.~\ref{fig:3}. Around the $\bar{K}$, $\bar{M}$, and $\bar{\Gamma}$ points, the vector $(d_1,d_2)$ forms a vortex, and as a result the energy bands [Figs.~\ref{fig:3}(a) and (b)] have Rashba splitting around $\bar{K}$, $\bar{M}$, and $\bar{\Gamma}$ points.
The spin distribution is in plane, and is shown in Fig.~\ref{fig:3}(c).
Usually, Rashba splitting appears around time-reversal-invariant momenta, 
where the spin degeneracy comes from the Kramers theorem. Nevertheless, in 
the present case, there is a Rashba splitting at the $\bar{K}$ point which is not invariant under time reversal. It is called peculiar Rashba splitting \cite{PhysRevLett.103.156801}
 which is different from the conventional Rashba splitting, because this peculiar Rashba splitting comes from the $C_{3z}$ symmetry at the $\bar{K}$ points which are not time-reversal invariant. 
These results qualitatively agree with the experimental results for $\beta$-Bi/Si(111) surface in Ref. \cite{PhysRevLett.103.156801}.

\section{Current-induced spin polarization in the Tl/Si surface}
As is already known, in systems with Rashba SOI, a charge current induces the spin polarization,
because an off-equilibrium electron distribution necessarily induces spin imbalance.
It was theoretically proposed in Refs.~\cite{Sov.Phys.JETP.73.537.1991,PhysRevLett.75.2004,PhysRevB.67.033104}
and experimentally observed in Refs.~\cite{PhysRevLett.93.176601,Nature.Phys.1.31}.
For the Rashba systems the total spin polarization is in-plane, and is perpendicular to the current in the plane.
Here we consider this current-induced spin polarization in systems with non-Rashba-type spin splitting, such as in the Tl/Si model.
Because the third term in the Tl/Si model in the Hamiltonian (\ref{eqn:1}) gives rise to the spins perpendicular to the crystal surface,
we can naively expect the current-induced 
spin polarization to have an out-of-plane component.

We numerically calculate the induced spin by the Boltzmann equation \cite{Mahan}.
We assume that the model
is uniform and that the impurity concentration is very
small.
In principle, the Boltzmann semiclassical equation can be obtained 
from the Wigner quantum kinetic equation as a semiclassical limit, and 
the classical distribution function is written as $f_{\sigma , \sigma '}(\vec{r},\vec{k},t)$,
which depends on spin indices.
In this paper, 
our goal is to qualitatively discuss various aspects of the non-Rashba SOC, 
and not to quantitatively calculate physical quantities in specific materials. 
Therefore, for simplicity
of calculation, we assume the classical distribution function to be 
spin independent, $f(\vec{r},\vec{k},t)$. It means that the relaxation time is 
assumed to be independent of spin. 
In addition, the classical distribution function $f$ does not depend on $\vec{r}$, because we assume this system to be uniform.
The Boltzmann equation becomes
\begin{eqnarray}
0=\frac{\partial \vec{k}}{\partial t}\cdot\vec{\nabla}_rf+\left(\frac{df}{dt}\right)_
{\mathrm{collisions}},\label{eqn:s-1}
\end{eqnarray}
where the last term is the temporal change rate by collisions with impurities.
The equation of motion for $\frac{\partial \vec{k}}{\partial t}$ in the absence of an external magnetic field is given by 
$\frac{\partial \vec{k}}{\partial t}=-e\vec{E}.$
We approximate the collisions term by relaxation-time approximation:
\begin{eqnarray}
\left(\frac{df}{dt}\right)_{\mathrm{collisions}}&=&-\frac{f(\vec{k})-f_0(\vec{k})}{\tau_t(\vec{k})},\label{eqn:s-3}
\end{eqnarray}
where $f_0(\vec{k})$ is the Fermi distribution function, and  $\tau_t(\vec{k})$ is the relaxation time.
Here, we assumed that the relaxation is due to impurity scattering, and is independent of spin, in order to illustrate the role of the non-Rashba SOC in the 
current-induced spin polarization in a simpler fashion.
In particular, we neglect scattering by phonons 
which may lead to anomalous Cherenkov effect due to the interaction between electrons and the crystal lattice with SOC \cite{PhysRevB.83.081308}, because we assume the electric field not too strong so that the electrons are subsonic, and, thus, the scattering rate on impurities may dominate over the Cherenkov dissipation.
By taking an approximation of replacing $f$ by $f_0$ in the first term of Eq.~(\ref{eqn:s-1}),
the distribution function under an electric field is written as
\begin{eqnarray}
f(\vec{k})&{\approx}&f_0(\vec{k})-e\tau_t(\vec{k})\vec{E}\cdot\vec{\nabla}_{\vec{k}}f_0(k).
\label{eqn:s-4}
\end{eqnarray}
The spin polarization is written as 
\begin{eqnarray}
{\langle}{S_i}{\rangle}=\frac{1}{2}\int\frac{d\vec{k}}{(2\pi)^2}{\langle}{s_i}{\rangle}\left(-e
\tau_t(\vec{k})\vec{E}\cdot \vec{\nabla}_{\vec{k}}f_0(\vec{k})\right),\label{eqn:4}
\end{eqnarray}
where ${\langle}s_i\rangle={\eta}\hat{d}_i$ is the spin expectation value.

\begin{figure}[td]
 \begin{center}
  \includegraphics[width=85mm]{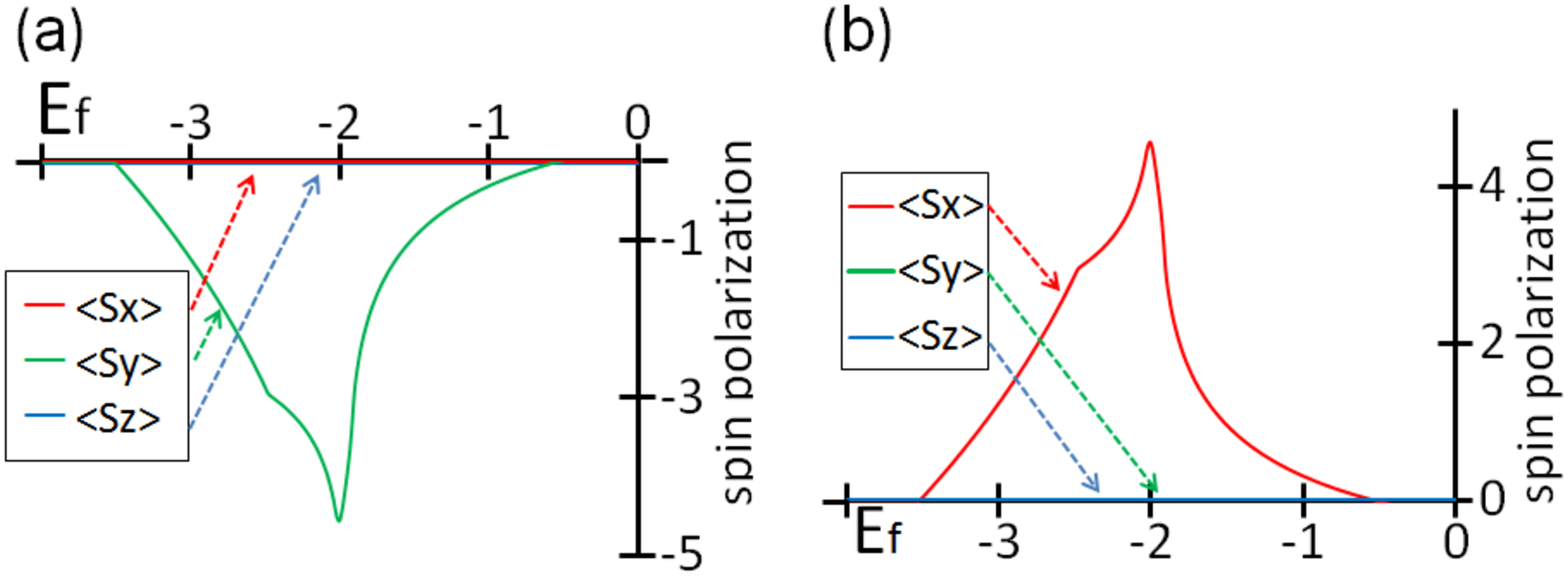}
 \end{center}
 \caption{(Color online)\ Spin polarization induced by the electric field along the (a) $x$ direction and (b) $y$ direction. Spin polarization is shown in a unit of $\frac{e{\tau}E_i}{2(2\pi)^2}$.}
 \label{fig:5}
\end{figure}

The numerical results are plotted in Fig.~\ref{fig:5}, where we take $t=1$ and $\lambda_x=\lambda_z=0.1$ and set $\tau_t$ to be a constant $\tau$.
Here we note that 
the nonzero spin polarization in Fig.~\ref{fig:5} arises from the spin expectation values for
spin-split bands, 
because the relaxation time is set as a constant, and 
the spin dependence is considered only through the spin expectation value.
It is also the case for the conventional spin-orbit-coupled systems such 
as Rashba systems \cite{Sov.Phys.JETP.73.537.1991,PhysRevLett.75.2004,PhysRevB.67.033104}.
The electric field is taken along the $x$ direction in Fig.~\ref{fig:5}(a) and along the $y$ direction in Fig.~\ref{fig:5}(b).
The result is shown 
as 
the dimensionless spin polarization which is the spin polarization divided by $\frac{e{\tau}E_i}{2(2\pi)^2}$ ($i=x,y$), as a function of the Fermi energy $E_f$. 
We see that for both cases the 
induced spins are along the direction which is the direction rotated by $-90^\circ$ from the electric field, 
which follows from symmetry as we show later.
From these results, in spite of the presence of the third term in Eq.~(\ref{eqn:2}) having the out-of-plane spins, the current does not induce out-of-plane spin polarization within linear response.
It means that the contributions from the $\bar{K}$ and $\bar{K'}$ points cancel each other.

\begin{figure}[htb]
 \begin{center}
  \includegraphics[width=85mm]{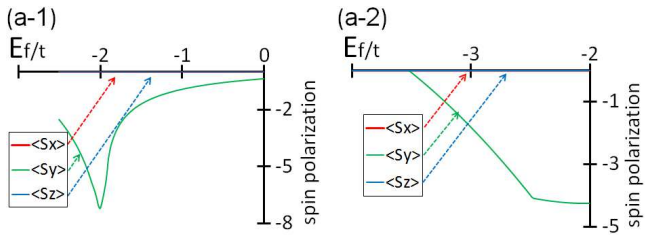}
 \end{center}
 \caption{(Color online)\ Spin polarization induced by the current for the states around (a-1) $\bar{M}$ and (a-2) $\bar{K}$ in response to $E_x$. Spin polarization is shown in a unit of $\frac{e{\tau}E_i}{2(2\pi)^2}$.}
 \label{fig:6}
\end{figure}

The results in Fig.~\ref{fig:5} have three characteristic energies: $E_f^{(1)}=-3.5$ where the spin polarization arises, $E_f^{(2)}=-2.3$ for the kink, and $E_f^{(3)}=-2$ for the peak.
From the energy band [Fig.~\ref{fig:1}], $E_f^{(1)}=-3.5$ and $E_f^{(2)}=-2.3$ correspond to band bottoms of the spin-split bands near the the $\bar{K}$ point (Fig.~\ref{fig:1}(a)).
At $E_f^{(3)}=-2$, there are band bottoms of spin-split bands at the $\bar{M}$ point (Fig.~\ref{fig:1}(b)).
To see the reasons for the characteristic values of the Fermi energy $E_f$ in Fig.~\ref{fig:1}, we calculate the contribution of the induced spin polarization, coming from the states near the $\bar{K}$ and that from the $\bar{M}$ points.
The results for the spin polarization are shown in Fig.~\ref{fig:6}.
We take the numerical calculation range of Fermi energy set as $-2\geq{E_f}$, because the approximate Hamiltonian is effective only in the vicinity of each point.

The contribution from the states near $\bar{M}$ point [Fig.~\ref{fig:6}(a-1)] has a peak at $E_f=-2$. Thus, the peak of the spin polarization in Fig.~\ref{fig:5} is due to the contribution from the states close to the $\bar{M}$ points.
The strong peak due to the states close to the $\bar{M}$ point might be attributed to a relatively flat dispersion around the $\bar{M}$ point, giving rise to a high density of states.
On the other hand, the contribution from the $\bar{K}$ point (Fig.~\ref{fig:6}(a-2)), has a kink at $E_f=-2.3$. 
The kink at $E_f^{(2)}$ is due to the contribution from the states close to the $\bar{K}$ points.


To see the reason for the absence of out-of-plane spin polarization in Fig.~\ref{fig:5}, we express the linear response of the spin polarization to the current as
\begin{eqnarray}
\langle S_i\rangle=\phi_{ij}J_j,\label{eqn:5}
\end{eqnarray}
where 
$\phi_{ij}$ represents response coefficients and $J_j\ (j=x,y)$ is the current.
By imposing crystal symmetries to Eq.~(\ref{eqn:5}), we obtain $\phi_{xy}=-\phi_{yx}$ and $\phi_{ij}=0$ otherwise. Thus, it is written as 
\begin{eqnarray}
\langle \vec{S}\rangle =\phi(\vec{J}\times\vec{z}),\label{eqn:6}
\end{eqnarray}
where $\phi$ is a real constant.
From this calculation, we find that the spin polarization is always in-plane, and is always perpendicular to the in-plane current.
It agrees with our numerical calculations and with the calculation by Liu {\it et al}. \cite{PhysRevB.80.241304} with four-band model.
Therefore, in the present case the crystal symmetries prohibit the out-of-plane spin polarization, despite the nonzero out-of-plane spins for the states near $\bar{K}$ and $\bar{K}'$ points.


\section{Persistent spin helix in the Tl/Si}

In the model of Tl/Si, the spin splitting around the $\bar{K}$ and $\bar{K'}$ points brings about an out-of-plane spin texture around these points, similar to thin films of transition metal dichalcogenides \cite{PhysRevLett.108.196802, nnano.2012.95, nnano.2012.96}.
In such cases, nesting between the two pockets at $\bar{K}$ and $\bar{K'}$ brings about an 
anomalously long lifetime of a spin excitation at the nesting wave vector, as proposed 
in the two-dimensional electron gas with equal size of the Rashba and Dresselhaus spin-orbit couplings \cite{PhysRevLett.97.236601}. The key ingredients of this novel phenomenon are the two nested Fermi surfaces by the magic shifting vector having the opposite spins. In addition, the enhanced spin lifetime has been confirmed experimentally in semiconductor heterostructures \cite{PhysRevLett.98.076604,N458610}.

\begin{figure}[htb]
 \begin{center}
  \includegraphics[width=85mm]{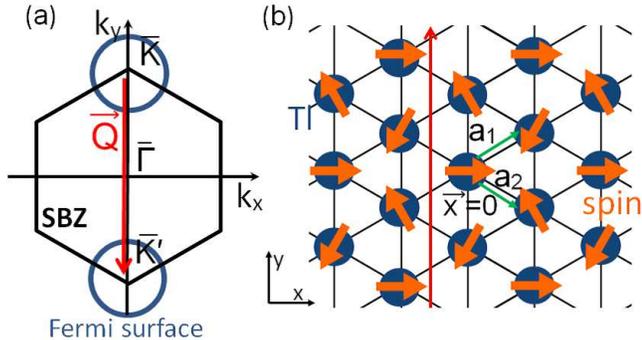}
 \end{center}
 \caption{(Color online)\ (a) The surface Brillouin zone (SBZ) in Tl/Si and the Fermi surfaces (blue circles) around $\bar{K}$ and $\bar{K'}$ points near $E_f=-2$. The red arrow indicates the magic shifting vector $\vec{Q}$. (b) Schematic illustration of the spin helix. The orange arrows indicate the spin on the Tl atoms (blue balls). The directions of the red arrows represent the in-plane spin directions forming the spin helix. The vectors of $\vec{a_1}$, $\vec{a_2}$ (green arrows) indicate primitive translation vectors.}
 \label{fig:8}
\end{figure}

We apply this theory to the Tl/Si model. We assume that the Fermi energy is controlled by doping, 
so that there are
small electron pockets around $\bar{K}$ and $\bar{K}'$. Then the system has the Fermi surfaces shifted by 
the magic shifting vector $\vec{Q}=[0,-8\pi/(3a)]$ from the $\bar{K}$ point to the $\bar{K'}$ point [Fig.~\ref{fig:8}(a)] and the spins have the opposite direction
perpendicular to the $xy$ plane at the $\bar{K}$ and $\bar{K'}$ points [Fig.~\ref{fig:1}(a)].
For example, let us take a state with the spin along the $x$-direction at $\vec{x}=\vec{0}$,
described by 
\begin{eqnarray}
|\psi\rangle _{\bar{\Gamma}} =e^{i\left(\vec{k_0}-\frac{\vec{Q}}{2}\right){\cdot}{\vec{x}}}\left(\begin{array}{c}1\\0\end{array}\right)+
\mathrm{e}^{i\left(\vec{k_0}+\frac{\vec{Q}}{2}\right){\cdot}\vec{x}}\left(\begin{array}{c}0\\1\end{array}\right),
\end{eqnarray}
where the $\vec{k}_0$ ($|\vec{k}_0|\ll \pi/a$) is a wave vector. It is a superposition of two
eigenstates at $\bar{K}$ and at $\bar{K}'$, which are degenerate. The spin expectation value at arbitrary $\vec{x}$ is written as
\begin{eqnarray}
{\langle}S_x{\rangle}&=&\frac{1}{2}\langle\psi|\sigma_x|\psi{\rangle}
={\cos}\left(\vec{Q}\cdot\vec{x}\right),\\
{\langle}S_y{\rangle}&=&\frac{1}{2}\langle\psi|\sigma_y|\psi{\rangle}
=-{\sin}\left(\vec{Q}\cdot\vec{x}\right),\\
{\langle}S_z{\rangle}&=&\frac{1}{2}\langle\psi|\sigma_z|\psi{\rangle}
=0.
\end{eqnarray}
Hence, the rotation angle of spin moving from $\vec{x}=\vec{0}$ to $\vec{x}\neq 0$ is $\vec{Q}\cdot \vec{x}$.
Therefore, the rotation angles of spin from $\vec{x}=\vec{0}$ to $\vec{x}=\vec{a_1}$ and $\vec{a_2}$ are
\begin{eqnarray}
\vec{Q}\cdot \vec{a_1}=\dfrac{4\pi}{3},\ \ \vec{Q}\cdot \vec{a_2}=-\dfrac{4\pi}{3}.
\end{eqnarray}
From this calculation we obtain Fig.~\ref{fig:8}(b) as a spin texture on the Tl atoms forming a triangular lattice.
It is a spin helix with the wave vector $\vec{Q}$. As is similar to the spin
helix in Ref.~\cite{PhysRevLett.97.236601}, it is expected to have enhanced spin lifetime, 
because of the degeneracy of the states around $\bar{K}$ and those around $\bar{K}'$. 
Therefore the spin helix in the Tl/Si model has in-plane spin texture with neighboring spins being different by $120^\circ$.
This is different from the spin helix in the 2D Rashba-Dresselhaus system \cite{PhysRevLett.97.236601},
where the spin helix is perpendicular to the surface and a value of the  magic shifting vector is much smaller than ours.
To realize the persistent spin helix in the Tl/Si system,  
hole doping \cite{Sakamoto647771} leads to emergence of two nested hole pockets with opposite spins at the $\bar{K}$ and $\bar{K}'$ points.
Once the spin texture is created, it will survive for a relatively long period.

\section{Bound state at a junction between two surface regions}

\begin{figure}[htb]
 \begin{center}
  \includegraphics[width=90mm]{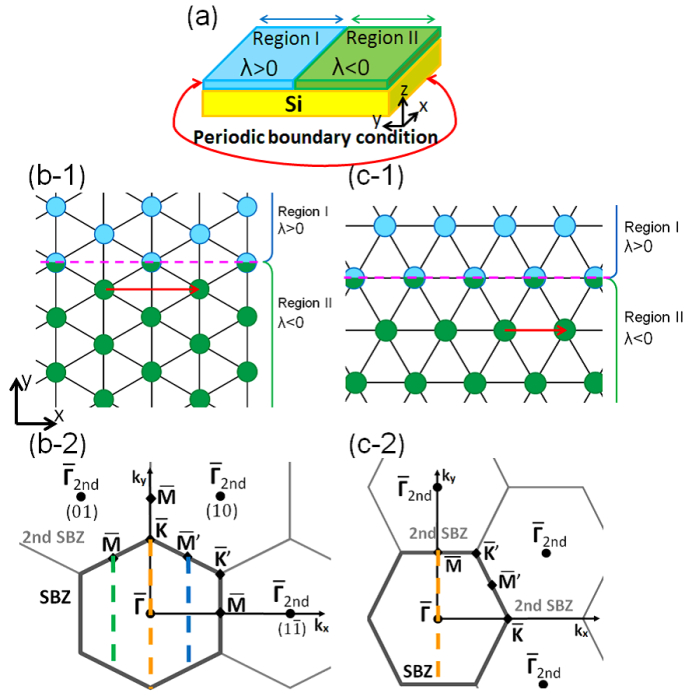}
 \end{center}
 \caption{(Color online)\ (a) Schematic illustration of the junction model. The blue (I) [green (II)] area indicates the region of the Bi or Tl layer where the SOI parameter has a negative (positive) value.
The yellow area indicates the Si substrate.
In (b) and (c), two choices of junctions are shown as a top view.
The primitive translation vectors (red arrow) are (b) $(\sqrt{3}a,0,0)$ and (c) $(a,0,0)$, respectively. The purple dashed line indicates the junction.}
 \label{fig:9}
\end{figure}

Next, we discuss bound states at a junction between two regions which have different signs of the SOI parameter, as in Fig.~\ref{fig:9}(a).
We consider the two models discussed in this paper, i.e., the Tl/Si system described by Eq.~(\ref{eqn:1}) without second term and the Bi/Si system described by Eq.~(\ref{eqn:3}). 
For the direction of the junction between two regions, we consider two cases shown in Figs.~\ref{fig:9}(b) and 7(c).
In the junction models, the SOI parameter ($\lambda_z$ in Tl/Si and $\lambda_y$ in Bi/Si)
is set as $+\lambda$ and $-\lambda$ in the regions I and II, respectively. To extract interface states, we compare the results with 2D bulk models, where the SOI parameter is set as $\lambda$ for the whole system.

\subsection{Bi/Si junction model}
\begin{figure}[h]
 \begin{center}
  \includegraphics[width=80mm]{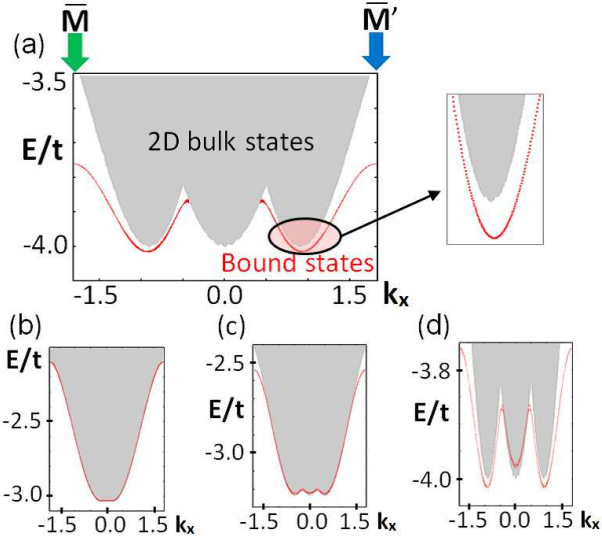}
 \end{center}
 \caption{(Color online)\ Energy bands of the 2D bulk states and the junction model for Bi/Si with different
SOI parameter such as (b) $|\lambda_y|=0.2t$, (c) $|\lambda_y|=0.5t$, and (a), (d) $|\lambda_y|=t$.   
(a) The red lines indicate the states when the SOI parameter is set as $+t$ in the blue region (I), $0$ on the junction boundary and $-t$ in the green region (II) (Fig.~\ref{fig:9}). The gray areas indicate the bulk bands with the SOI parameter set as $+t$.}
 \label{fig:10}
\end{figure}

We calculated the band structures for two types of Bi/Si junction models  shown in Figs.~\ref{fig:9}(b) and 7(c).
For Fig.~\ref{fig:9}(c), we could not find bound states at the junction. 
In contrast, for Fig.~\ref{fig:9}(b), we obtain the result in 
Fig.~\ref{fig:10}(a), which shows that 
there are bound states (red lines) below all the 2D bulk states (gray area) 
in Fig.~\ref{fig:10}(a).
The decay length depends on the SOI parameter; for example it is $10.0$ times the lattice spacing for the parameters $|\lambda_y|=t$.
Additionally, we find that the energy difference between the bound states and the bulk band edge is larger for a larger value of the SOI parameter, as seen from the results with the SOI parameter $|\lambda_y|=0.2t$, $|\lambda_y|=0.5t$, and $|\lambda_y|=t$ [Figs.~\ref{fig:10}(b)-8(d)).

Furthermore, in Fig.~\ref{fig:10}(a), there are four bound states forming two Kramers pairs. The Kramers degeneracy appears because the system restores spatial inversion symmetry when the SOI parameter are set as $-\lambda_y(y<0)=\lambda_y(y>0)$.
In contrast, when the SOI parameter is set as $-\lambda_y(y<0)\neq \lambda_y(y>0)$,
the degeneracy of the bound states are lifted at $\vec{k}$ points other than the $\bar{M}$ points.

Additionally, spin directions of these bound states are almost along the $z$ axis, with a small value of $y$ component.
This is expected from the crystal symmetries. Because this Bi/Si junction model [Fig.~\ref{fig:9}(b)] has $M_{yz}$ and the time-reversal symmetries, the $x$ component of spin necessarily vanishes. 
The out-of-plane spin directions of the bound states are in sharp contrast to the in-plane spin directions of the bulk states. 
We note that these interesting bound states cannot be reproduced by continuum models as we discuss later.

Next, we try to analytically calculate bound states for junction systems in order to compare with the above numerical results.
First, we note that in the Bi/Si junction model, the bound state of Fig.~\ref{fig:10}(a) appears around the $\bar{M}$ or $\bar{M'}$ points. These points are shown as 
green and blue arrows in Fig.~\ref{fig:10}(a), corresponding to the dotted lines in Fig.~\ref{fig:9}(a) and at these points 
the spin splitting is of the non-Rashba type. Therefore, we take the Hamiltonian from an expansion of Eq.~(\ref{eqn:3}) around the $\bar{M}$ point:
\begin{eqnarray}
H\left( k \right) = \frac{k^2}{2m}+\lambda _y \left( \sigma_xk_y+\sigma_yk_x\right).\label{eqn:7}
\end{eqnarray}
We then consider a junction between two surface regions, described by the Hamiltonian (\ref{eqn:7})
with the SOI parameter changing along the $y$ direction $\lambda _y =\lambda (y)$.
We tried the calculation of the bound states of this model, with two different choices of the function $\lambda (y)$: the step function and the hyperbolic tangent function, as explained in the
Appendix.
As a result, the continuum model in the Bi/Si junction model cannot reproduce the bound states.
The reason for this is yet to be clarified, and left as a future work.
We note that for an interface between two regions of the one-dimensional nanowire with different directions of the SOI vector,
bound states are found analytically \cite{JelenaKlinovaja},
whereas the system considered is different from ours. 

To realize this junction system, we note that the SOI parameter in this system originates from the broken inversion symmetry due to the surface, similar to the Rashba SOI.
Therefore, realization of our junction model with 
different signs of SOI parameters would be similar to the case for the Rashba SOI.
Here, we note an example of a related system, a 
non-centro-symmetric semiconductor BiTeI. In BiTeI, the stacking order of atomic layers, Bi-Te-I or I-Te-Bi determines the sign of the Rashba parameter in this system, and by changing
the stacking order it is possible to achieve a different sign of Rashba SOI \cite{PhysRevB.89.085402}. 
Although the present system is different from BiTeI, it might be possible to 
achieve a junction between two regions with different signs of SOI parameters
by controlling the atomic arrangement in each region. 


\subsection{Tl/Si junction model}
\begin{figure}[h]
 \begin{center}
  \includegraphics[width=80mm]{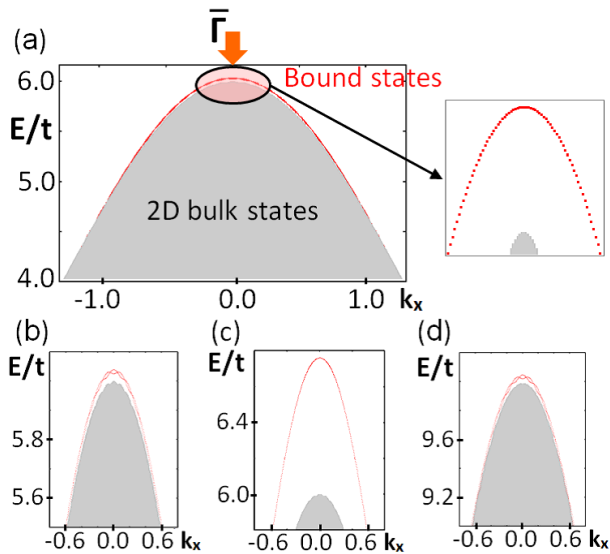}
 \end{center}
 \caption{(Color online)\ Energy bands of the 2D bulk states and the junction model for Tl/Si with different
SOI parameter such as (a), (b) $|\lambda_z|=t$, (c) $|\lambda_z|=1.73t$, and (d) $|\lambda_z|=2.5t$.   
(a) The red lines indicate the states when the SOI parameter is set as $+t$ in the blue region (I), $0$ in the junction boundary, and $-t$ in the green region (II) (Fig.~\ref{fig:9}). The gray areas indicate the bulk bands with the SOI parameter set as $+t$.}
 \label{fig:11}
\end{figure}

Next, we numerically analyze two types of Tl/Si junction models shown in Figs.~\ref{fig:9}(b) and 7(c).
There are no bound states in the Tl/Si junction model of Fig.~\ref{fig:9}(c).
Meanwhile, in the Tl/Si junction model of Fig.~\ref{fig:9}(b)
there appear bound states (red lines) above all the 2D bulk states (gray area) 
in Fig.~\ref{fig:11}(a).
The results for various values of the SOI parameter are shown in Figs.~\ref{fig:11}(b)-9(d).
From these results, 
the energy difference between the bound states and the bulk band edge is non-monotonic and it is maximized at about $|\lambda_z|=1.73t$ [Fig.~\ref{fig:11}(c)] in our case.
The decay length of the bound states also depends on the SOI parameter, 
and it is $38.5$ times the lattice spacing, for the parameters $|\lambda_z|=t$.

As is similar to the Bi/Si junction in the previous section,  
the four bound states in Fig.~\ref{fig:11}(a) form two Kramers pairs, 
stemming from the restored spatial inversion symmetry when $-\lambda_z(y<0)=\lambda_z(y>0)$ holds.
This degeneracy at $\vec{k}$ points other than the $\bar{\Gamma}$ point is
lifted when $-\lambda_z(y<0)\neq\lambda_z(y>0)$.
%
We also note that the spin of the bound states is along the $z$ axis because the effective Hamiltonian (Eq.~(\ref{eqn:1})) without the second term only has the $\sigma_z$ term.

Next, we construct a continuum model for the present system 
and calculate bound states for the junction system in
order to compare with the numerical results of the tight-binding models.
The bound states for Fig.~\ref{fig:9}(b) at $k_x=0$ [orange arrow at Fig.~\ref{fig:11}(a)] should correspond to
$\bar{K}$, $\bar{M}$, or $\bar{\Gamma}$ points [orange dotted line at Fig.~\ref{fig:9}(b-2)], whereas in Fig.~\ref{fig:9}(c) at $k_x=0$ corresponding to $\bar{M}$ and $\bar{\Gamma}$ points there are no bound states [orange dotted line at Fig.~\ref{fig:9}(c-2)].
Therefore, the bound states of Fig.~\ref{fig:11}(a) are expected to come from the $\bar{K}$ point. For this reason, by expanding Eq.~(\ref{eqn:1}) around the $\bar{K}$ point we take the following continuum Hamiltonian:
\begin{eqnarray}
H\left( k \right) = \frac{k^2}{2m}+\lambda _z \sigma _z \left( k_y^2+k_x^2 \right).\label{eqn:8}
\end{eqnarray}
We take the same procedure for calculation of the bound states for this continuum model [Eq.~(\ref{eqn:8})]
in the Appendix. As a result, the continuum model cannot reproduce the bound states.
Thus, we conclude that the bound states found in the junction system 
are unique to the lattice model.


\section{Conclusion}
In conclusion, we derive effective nearest-neighbor tight-binding Hamiltonians for surfaces of Tl/Si and Bi/Si by taking into account the crystal symmetries. The energy band of the Tl/Si model has non-Rashba splitting and the spin direction for each band is perpendicular to the crystal surface at the $\bar{K}$ point.
The energy band of the Bi/Si model has peculiar Rashba splitting at the $\bar{K}$ point, because the Rashba splitting occurs around the $\bar{K}$ points which are not 
time-reversal invariant.
Our results of the model calculation qualitatively agree with the experimental results.
In addition, differences between the Tl/Si model and the Bi/Si model lie in the out-of-plane spin at the $\bar{K}$ points in the Tl/Si model. This difference originates from the non-Rashba term involving the out-of-plane spins in Eq.~(\ref{eqn:1}). 

Additionally, as one of our theoretical explorations towards novel spin properties, we calculate current-induced spin polarization for the Tl/Si model.
Although the third term in the Tl/Si model in the Hamiltonian (Eq.~(\ref{eqn:1})) has a spin component perpendicular to the crystal surface, the out-of-plane component of current-induced spin polarization within linear response is zero, as prohibited by 
the crystal symmetries.
As a result the current-induced spin polarization of the non-Rashba model is in-plane, an d perpendicular to the current, which is qualitatively similar to that of Rashba model.

Furthermore, we apply the theory of persistent spin helix to the Tl/Si model. We find that in 
the spin helix in the Tl/Si model, the spins are shifted by $120^\circ$ between the neighboring sites and are in-plane, which is different from the spin helix in semiconductor heterostructure.

Finally, we numerically find bound states at the junctions between the two surface regions which have different signs of the SOI parameters in the Bi/Si system and in the Tl/Si system.
For the junction shown in Fig.~\ref{fig:9}(b) we found bound states at the junction.
From the lattice model we find that the spins of the bound states in the Bi/Si model is out of plane, which is in contrast with in-plane spin distribution in the bulk.

To summarize the whole results,  surface states allow various types of non-Rashba-type SOI due to low symmetries, and they are expected to lead to new spin properties which are absent in conventional Rashba systems, and they have much room for future research.

\begin{acknowledgments}
This work was supported by CREST, Japan Science and Technology Agency, and by Grant-in-Aid for Scientific Research on Innovative Areas (No.~26103006), Grant-in-Aid for Challenging Exploratory Research (No.~26600012) from the Ministry of Education, Culture, Sports, Science and Technology (MEXT) of Japan.

\end{acknowledgments}


\appendix*
\section{}

In this appendix, we analytically calculate the bound states for the Bi/Si junction model of Fig.~\ref{fig:9}(b).
First, we consider a junction between two surface regions described by the Hamiltonian (\ref{eqn:7}) with different signs of the SOI parameter, $\lambda_y(y) \rightarrow \pm \lambda_0$ for $y\rightarrow \pm \infty$.
Several choices of $\lambda_y(y)$ are possible.
First, we take $\lambda_y(y)$ to have the form of the step function along the $y$ direction $\lambda_y(y)=\lambda_0\theta (y)$, where $\theta(y)$ is step function.
The system is set to be infinite along the $x$ direction.
We should replace $k_y$ by $-i\partial_y$ because there is no translational symmetry along the $y$ axis. 
Furthermore, we replace $\lambda_y(y)[\sigma_x(-i\partial_y)]$ by $-i\frac{1}{2}\sigma_x[\lambda_y(y)\partial_y+\partial_y\lambda_y(y)]$ to preserve Hermiticity of the Hamiltonian. 

Along the $y$ direction, we call the two regions I ($y>0$) and I\hspace{-1pt}I ($y<0$).
From the Hamiltonian rewritten from Eq.~(\ref{eqn:7}), boundary conditions are written as
\begin{eqnarray}
\psi_I (0) &=& \psi_{I\hspace{-1pt}I}(0), \label{eqn:A-1}\\
\dfrac{1}{2m}\left\{ \partial_y\psi_I(0)-\partial_y\psi_{I\hspace{-1pt}I}(0)\right\}&=&-i\sigma_x \lambda_0 \psi_I(0).\label{eqn:A-2}
\end{eqnarray}
Meanwhile, when the wave function of the bound states decays exponentially as $e^{\pm \kappa_{\pm}y}$,
the eigen energy is
\begin{eqnarray}
E_{\pm}&=& \dfrac{1}{2m}\left( k_x^2-\kappa_{\pm} ^2\right)\pm \lambda_0\sqrt{k_x^2-\kappa_{\pm} ^2},\nonumber \\
&=&\dfrac{C_{\pm}^2}{2m}\pm \lambda_0 C_{\pm},\label{eqn:A-3}
\end{eqnarray}
where $C_{\pm}\equiv \sqrt{k_x^2-\kappa_{\pm}^2}$.
Here, because the values of $\lambda_y(y)$ for the regions I and I\hspace{-1pt}I have the same size with different signs, the decay of the wave functions in region I\hspace{-1pt}I is characterized by $\kappa_{\pm}$, as is the same with region I.
Therefore, $E_{I\pm}=E_{I\hspace{-1pt}I\mp}$ follows, and the bound states of two regions are written as 
\begin{eqnarray}
\psi_{I}=&\alpha & \left( \begin{array}{c}i(\kappa_{+} -k_x)\\C_{+}\end{array}\right)e^{-\kappa_{+}y+ik_xx}\nonumber\\
&+&\beta \left( \begin{array}{c}i(\kappa_{-} -k_x)\\ - C_{-}\end{array}\right)e^{-\kappa_{-}y+ik_xx},\ \label{eqn:A-4}
\end{eqnarray}
\begin{eqnarray}
\psi_{II}=&\alpha{'}& \left( \begin{array}{c}i(\kappa_{+}+k_x)\\C_{+}\end{array}\right)e^{\kappa_{+}y+ik_xx}\nonumber\\
&+&\beta{'} \left( \begin{array}{c}i(\kappa_{-}+k_x)\\ - C_{-}\end{array}\right)e^{\kappa_{-}y+ik_xx},\ \label{eqn:A-5}
\end{eqnarray}
where $\alpha$, $\beta$, $\alpha '$ and $\beta '$ are constants.
We then impose the boundary conditions to derive $\alpha$, $\beta$, $\alpha '$ and $\beta '$ of Eqs.\ (\ref{eqn:A-4}) and (\ref{eqn:A-5}). As a result, we found that there are no such values which satisfy the boundary conditions.

\begin{figure}[h]
 \begin{center}
  \includegraphics[width=85mm]{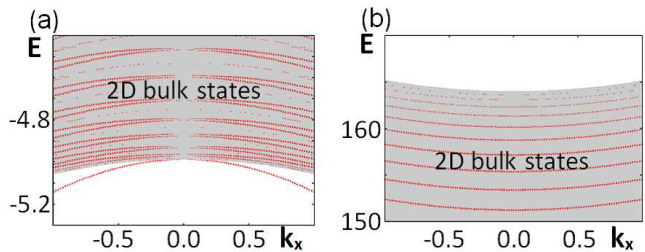}
 \end{center}
 \caption{(Color online)\ Energy band of the 2D bulk states and the junction model for (a) Bi/Si and (b) Tl/Si. The red lines indicate the states when the SOI parameter is set as $\lambda_0\tanh(\frac{y}{a})$. The gray area indicates 2D bulk states when the SOI parameter is set as $\lambda_0$ in all the regions. The other parameters are set as $\lambda_0=40$, $m=0.5$, and $a=0.001$.}
 \label{fig:13}
\end{figure}

Second, we try with the SOC parameter $\lambda_y(y)$ to have the form $\lambda_y(y)=\lambda_0\tanh(\frac{y}{a})$, where $a$ represents a width of the domain wall.
The results of the Bi/Si junction model are drawn as red lines and the bulk band is drawn as a gray area in Fig.~\ref{fig:13}(a). According to Fig.~\ref{fig:13}(a), there are no states below all 2D bulk states around $\bar{M}$ point ($k_x=0$). Altogether, we cannot reproduce the bound states by this continuum model.

Next, we analytically calculate the bound states for Tl/Si junction model of Fig.~\ref{fig:10}(b). We consider a junction between two surface regions described by Hamiltonian Eq.~(\ref{eqn:8}) with different signs of the SOI parameter along the $y$ direction. We try with the SOI parameter $\lambda_z$ to have the form $\lambda_z=\lambda_0\tanh(\frac{y}{a})$. We take the calculation procedure used for the Tl/Si junction model and obtain the energy band in Fig.~\ref{fig:13}(b).
According to Fig.~\ref{fig:13}(b), there are no states above all 2D bulk states around $\bar{K}$ point ($k_x=0$). Altogether, we cannot reproduce the bound states by this continuum model.

\end{document}